\documentclass[preprintnumbers,a4paper,twocolumn,floats,aps,nofootinbib]{revtex5}
\usepackage{graphicx}
\usepackage{amsmath,amssymb,amsfonts}
\usepackage{hyperref}
\usepackage{bbm}
\usepackage{xspace}
\newcommand{\eVdist}{\kern-0.06em}

\newcommand{\mev}{\:\text{Me\eVdist V}}
\newcommand{\gev}{\:\text{Ge\eVdist V}}

\def \beq{\begin{equation}}
\def \eeq{\end{equation}}
\def \bea{\begin{eqnarray}}
\def \eea{\end{eqnarray}}

\newcommand{\D}{\mathrm{d}}

\newcommand{\MO}{{\tt MicrOmegas}\xspace}

\newcommand{\CalcHep}{{\tt CalcHep}\xspace}
\newcommand{\PY}{{\tt PYTHIA}\xspace}

\begin{document}
\title{Constraints on light mediators: confronting dark matter searches with $B$ physics}

\preprint{CERN-PH-TH/2013-250, Bonn-TH-2013-22, DESY~13-192}

\author{Kai Schmidt-Hoberg$^a$}
\email{kai.schmidt-hoberg@cern.ch}
\author{Florian Staub$^b$}
\email{fnstaub@th.physik.uni-bonn.de}
\author{Martin Wolfgang Winkler$^c$}
\email{martin.winkler@desy.de}

\affiliation{$^a$Theory Division, CERN, 1211 Geneva 23, Switzerland}
\affiliation{$^b$Bethe Center for Theoretical Physics \& Physikalisches Institut der 
Universit\"at Bonn, \\
Nu{\ss}allee 12, 53115 Bonn, Germany}
\affiliation{$^c$Deutsches Elektronen-Synchrotron DESY, 
Notkestra\ss e 85, D-22607 Hamburg, Germany}

\begin{abstract} 
Light scalars appear in many well-motivated extensions of the Standard Model including supersymmetric models with additional gauge singlets. Such scalars could mediate the interactions between dark matter and nuclei, giving rise to the tentative signals observed by several dark matter direct detection experiments including CDMS-Si. In this letter, we derive strong new limits on light scalar mediators by using the LHCb, Belle and BaBar searches for rare $\Upsilon$ and $B$ decays. These limits rule out significant parts of the parameter space favored by CDMS-Si. Nevertheless, as current searches are not optimized for investigating weakly coupled light scalars, a further increase in experimental sensitivity could be achieved
by relaxing requirements in the event selection.
\end{abstract}

\maketitle

\section{Introduction}

Recent results in dark matter (DM) direct detection, notably the excesses over the expected background seen by CDMS-Si~\cite{Agnese:2013rvf},
CoGeNT~\cite{Aalseth:2011wp} and CRESST-II~\cite{Angloher:2011uu} as well as the long-standing annual modulation signal seen
by DAMA~\cite{Bernabei:2010mq} have been interpreted as relatively light DM with mass $m_{\chi} \lesssim 10 \gev$.
While the cross sections inferred from DAMA and CRESST-II are in strong tension with null results from other leading experiments such as
XENON~\cite{Angle:2011th,Aprile:2012nq}, the CDMS-Si signal can be (marginally) consistent with all other searches even under the standard 
assumption of elastic spin-independent scattering~\cite{Frandsen:2013cna}. This makes an interpretation in the context of simple DM models particularly attractive.
In this letter, we will concentrate on the case where a scalar field
mediates the interactions between DM and nuclei.
Collider searches have severely restricted the available model parameter space.
Very heavy scalar mediators are basically ruled out for the cross sections of interest due to monojet searches, even for the case of Higgs-like couplings~\cite{Haisch:2012kf,Cotta:2013jna}. 
Light mediators, however, are much harder to constrain as they increase the relative sensitivity of direct DM searches with respect to colliders.

In this letter, we study bounds on scalars with reduced Higgs-like couplings to SM states which are lighter than the DM.
In particular we derive constraints coming from $\Upsilon$ and $B$ meson decays at Belle, BaBar and LHCb. While these bounds apply
generally to light scalars coupled to the SM, they are particularly interesting when the scalar mediates interactions with the DM.
In this case, the bound on the coupling of the scalar to SM states can be rephrased as a bound on the DM direct detection cross section.
We find that the resulting bounds are very stringent and rule out much of the remaining parameter space which was considered viable
as an explanation for the recent experimental hints in DM searches.

\section{A generic model}
For the purpose of this letter we study a model with light (Majorana) fermionic DM $\chi$ which is coupled to the SM via
a light (CP-even) scalar $\phi$, 
\begin{equation}
 \mathcal{L} = \mathcal{L}_\text{SM} - \frac{y\, m_f}{v}\,\phi\,\bar{f} f\ - \frac{1}{2}\kappa \phi \bar{\chi} \chi \; ,
\end{equation}
where $v\simeq 246\gev$ denotes the electroweak vev. 
Here we assumed that the couplings of $\phi$ to the SM fermions $f$ scale with their mass $m_f$ which is the case if the couplings are induced via the Higgs portal~\cite{Patt:2006fw}.\footnote{To obtain the Higgs portal induced couplings of $\phi$ to SM fields in the given form, one has to integrate out the Higgs boson~\cite{Batell:2009jf}.} This simple model is a natural limit of singlet extensions~\cite{Kappl:2010qx} of the Minimal Supersymmetric Standard Model (MSSM), where $\phi$ and $\chi$ stem from the same supermultiplet.\footnote{Such general singlet extensions of the MSSM have been motivated in~\cite{Lee:2011dya,Ross:2011xv,Ross:2012nr}.}
As far as the experimental constraints on the scalar $\phi$ are concerned, only its interactions with the SM are relevant, while its couplings
to DM can be ignored as long as the decay of $\phi$ into DM is kinematically forbidden. 
However, it is interesting to assume that this scalar mediates the predominant interactions between the visible and the dark sector as in our model.
In this case the coupling $\kappa$ can be inferred from requiring the correct relic density of the DM, $\Omega_\chi h^2 = 0.1199$~\cite{Ade:2013zuv}.
Then, the limit on the coupling $y$ as a function of the mediator mass $m_\phi$ can be rephrased as a limit on the DM direct detection cross section.

\begin{figure}[t]
\begin{center}  
  \includegraphics[width=6.5cm]{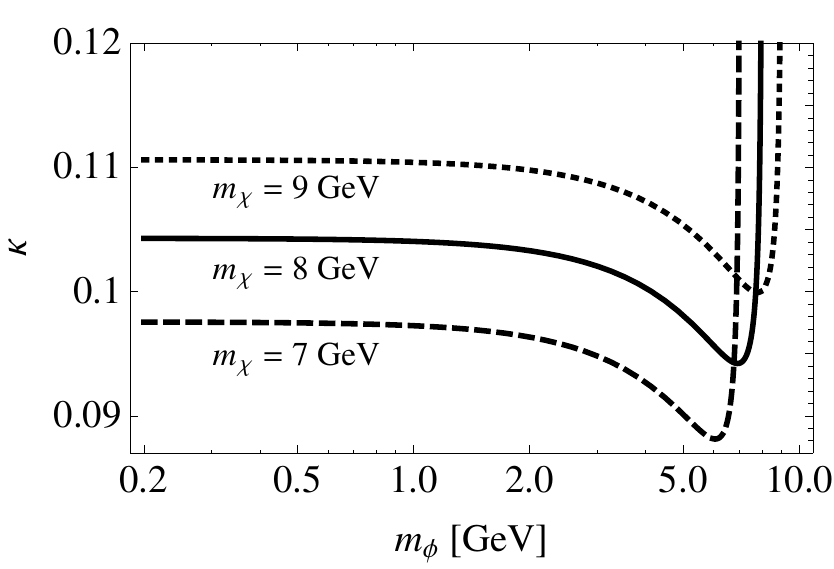}
\end{center}
\caption{Relic density as a function of the coupling $\kappa$ and the mediator mass $m_\phi$ for $m_{\chi} = 7,8,9\gev$.}
\label{fig:relickappa}
\end{figure}

For the case we study here, $m_\phi<m_\chi$, the annihilation of DM in the early universe is typically dominated by the process $\chi\chi\rightarrow\phi\phi$. 
The corresponding annihilation cross section can be estimated as $\sigma\, v_\mathrm{rel}   \simeq   \sigma_1  \, v_{\mathrm{rel}}^2$ with~\cite{Kappl:2010qx,Winkler:2012xwa}
\begin{align}
\label{eq:vrelexpansion}
 \sigma_1 & = \frac{\kappa^4m_{\chi}}{24\,\pi}  \, \sqrt{m_{\chi}^2-m_\phi^2} \; \frac{9 m_{\chi}^4-8 m_{\chi}^2 m_\phi^2+2m_\phi^2}{(2m_{\chi}^2-m_\phi^2 )^4} 
\end{align}
and $v_{\mathrm{rel}}$ denoting the relative velocity. For $p$-wave suppressed annihilations and $m_\chi = 5-10 \gev$ the DM relic density is achieved for $\sigma_1 \simeq 1.6 \cdot 10^{-25} \text{cm}^3/\text{s}$ (see e.g.~\cite{Winkler:2012xwa}). We have cross checked the analytic approximation by implementing the model into \CalcHep~\cite{Pukhov:2004ca} and using \MO~\cite{Belanger:2006is,Belanger:2007zz,Belanger:2010pz} for the relic density calculation. We find excellent agreement. The coupling $\kappa$ which yields the correct relic density for a given mass $m_\phi$ is shown in Fig.~\ref{fig:relickappa}.
The inferred value of $\kappa$ is rather insensitive to $m_\phi$ and the precise value of $m_\chi$. Except in the case of strong phase space suppression, we find $\kappa\simeq 0.1$.
Note that due to the $p$-wave nature of the annihilation, the model automatically evades the strong constraints on light DM arising from indirect detection (see e.g.~\cite{Kappl:2011kz,Ackermann:2011wa,Kappl:2011jw,Hooper:2012sr}).

The (spin-independent) DM nucleon cross section is dominated by $\phi$ exchange and can be written as
\begin{align}
 \sigma_{n}& = \frac{4 \mu_\chi^2}{\pi}\left(\frac{y \kappa}{2 v m_\phi^2}\right)^2 m_n^2\,(f^n_u+f^n_d+f^n_s+6/27 f_G)^2 \nonumber \\
  &  \simeq 10^{-40} \text{cm}^2 \left(\frac{\kappa}{0.1}\right)^2 \left(\frac{y}{0.01}\right)^2
 \left(\frac{\text{GeV}}{m_{\phi}}\right)^4 \;,
\end{align}
with $m_n$ the nucleon mass and $\mu_\chi$ the reduced mass of the DM-nucleon system. $f^n_{u,d,s}$ and $f_G$ are the scalar coefficients for the quark and gluon content of the nucleon which we take from~\cite{Belanger:2013oya}.\footnote{Note that in the given model, the cross sections for scattering off protons and neutrons are virtually indistinguishable.} This has to be compared with the cross section $\sigma_n\simeq 10^{-42}-10^{-40}\:\text{cm}^2$ suggested by the CDMS-Si result. If we take $\kappa$ to be fixed by the relic density requirement, the cross section only depends on the coupling $y$ for a given mediator mass $m_\phi$. In the next section we therefore study constraints on this coupling from rare decays.

\section{Constraints on light mediators from $\Upsilon$ and $B$ meson decays}
\begin{figure}[t]
\begin{center}  
  \includegraphics[height=4.4cm]{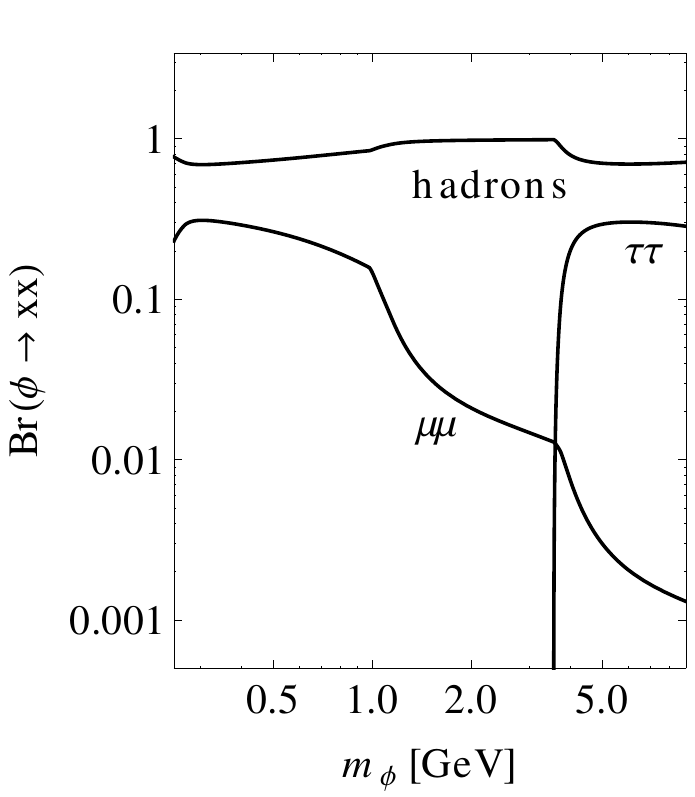}\hspace{4mm}
  \includegraphics[height=4.4cm]{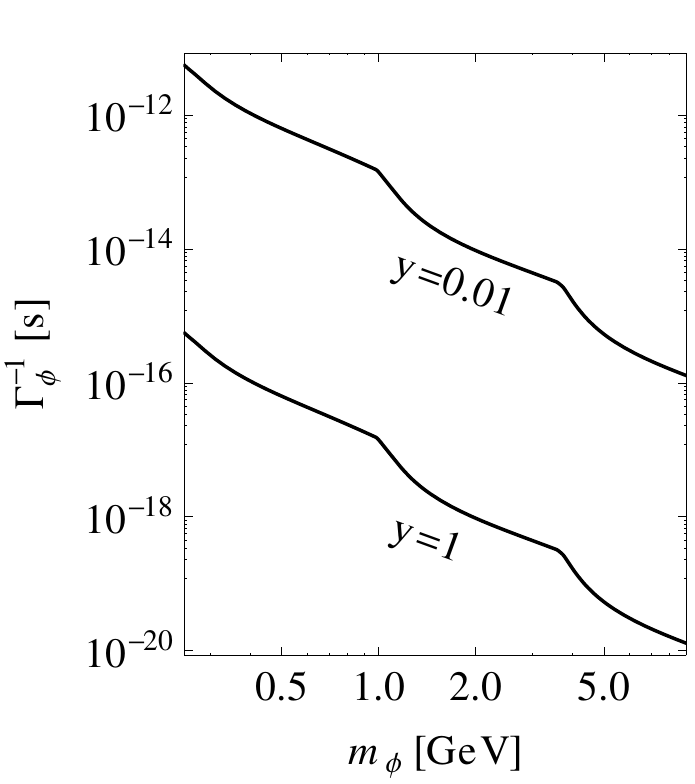}
\end{center}
\caption{Branching fraction of the light scalar $\phi$ into muons, taus and hadronic final states (left panel). Lifetime of $\phi$ for two different values of the coupling $y$ (right panel).}
\label{fig:branching}
\end{figure}

For comparison with experimental results we have to determine the decay pattern of $\phi$. The decay rate into muons is given as
\begin{equation}\label{eq:mudecay}
 \Gamma (\phi\rightarrow \mu\mu) \equiv \Gamma_{\mu\mu} = \frac{G_F\,m_\phi}{4\sqrt{2}\,\pi}\,y^2 m_\mu^2\,\beta_\mu^3\;,
\end{equation}
where $G_F$ is the Fermi constant and we defined $\beta_\mu = \sqrt{1-4m_\mu^2/m_\phi^2}\;\Theta(m_\phi-2\,m_\mu)$ with the Heaviside function accounting for the kinematical threshold of the decay mode. In order to determine the branching fraction into hadronic final states, we follow~\cite{McKeen:2008gd} and use the perturbative spectator model matched to chiral perturbation theory at the QCD scale. This leads to the following relative decay rates into muons, $\pi$-, $K$-, $\eta$-, $D$-mesons, taus and gluons
\begin{align}\label{eq:branching}
  &\: \Gamma_{\mu\mu}\: : \:\Gamma_{\pi\pi}\: : \:\Gamma_{KK}\: : \:\Gamma_{\eta\eta}\: : \:\Gamma_{DD}\: : \:\Gamma_{\tau\tau}\: : \:\Gamma_{gg}\nonumber \\
= & \: m_\mu^2\,\beta_\mu^3 \: : \: 3(m_u^2+m_d^2)\beta_\pi^3 \: : \:3\frac{9}{13}m_s^2\beta_K^3\: : \:
3\frac{4}{13}m_s^2\beta_\eta^3 \: : \nonumber \\ &  \: m_c^2\beta_D^3 \: : \: m_\tau^2\,\beta_\tau^3 \: : \: 
\left(\frac{\alpha _s m_\phi}{3 \pi}\right)^2 (6-2\beta_\pi^3-\beta_K^3)\;.
\end{align}
The effective light quark masses as well as $\alpha_s$ are determined from the matching which yields $m_u=m_d=50\mev$, $m_s=450\mev$ and $\alpha_s=0.47$~\cite{McKeen:2008gd}. In the left panel of Fig.~\ref{fig:branching}, we depict the resulting branching ratio into muons, taus and the summed branching into hadronic final states.\footnote{Our estimate of the branching fractions does not include bound state effects. We expect some deviations if $m_\phi$ resides in the vicinity of a scalar resonance.} The lifetime of the scalar $\Gamma_\phi^{-1}=(\Gamma_{\mu\mu}+\Gamma_{\pi\pi}+\dots)^{-1}$ is shown in the right panel of Fig.~\ref{fig:branching}. Note that for $\Gamma_\phi^{-1} \gtrsim 10^{-12}\: \text{s}$ the decay length of $\phi$ becomes relevant on detector scales.

\subsection{$\Upsilon$ decays}

\begin{figure}[t]
\begin{center}   
 \includegraphics[height=3.0cm]{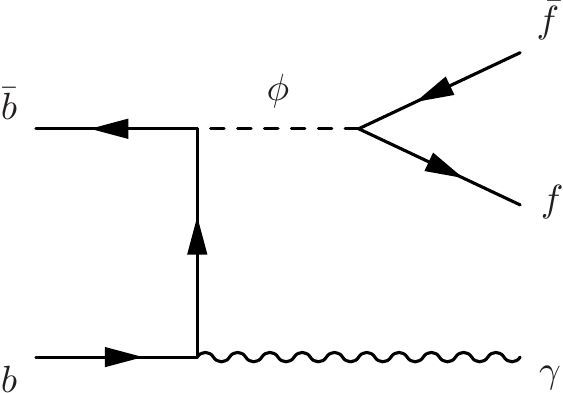}
\hfill
 \includegraphics[height=3.0cm]{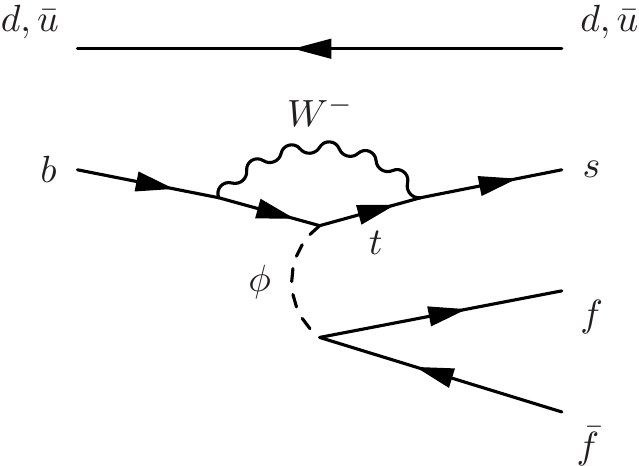}
\end{center}
\caption{Rare decays of $\Upsilon$ (left) and $B$ mesons (right) mediated by the light scalar $\phi$.}
\label{fig:penguin}
\end{figure}

If $m_\phi\lesssim 10\gev$, the light scalar can mediate the radiative decay $\Upsilon \rightarrow \gamma \, \phi$ with $\phi$ decaying further into mesons or leptons~\cite{Wilczek:1977zn} (see left panel of Fig.~\ref{fig:penguin}). In order to factor out uncertainties, it is reasonable to express the corresponding branching ratio in the form
\begin{equation}\label{eq:upsilon}
 \frac{\text{Br}(\Upsilon(nS)\rightarrow \gamma\,\phi)}{\text{Br}(\Upsilon(nS)\rightarrow e e )}=\frac{y^2 G_F m_b^2}{\sqrt{2}\pi\alpha}\,\mathcal{F}\,\Big(
1-\frac{m_\phi^2}{m^2_\Upsilon(nS)}\Big)\;,
\end{equation}
where $\alpha$ is the Sommerfeld constant, $m_b$ is the bottom mass and $\mathcal{F}$ a correction function which includes higher order QCD processes~\cite{Vysotsky:1980cz,Nason:1986tr} as well as bound state effects appearing when $m_\phi$ approaches the kinematical endpoint~\cite{Haber:1978jt,Ellis:1979jy}. A parameterization of $\mathcal{F}$ which includes both effects without double counting can be extracted from Fig.~1 in~\cite{Ellis:1985yb}.\footnote{Here we use the estimate (B) from Fig.~1 in~\cite{Ellis:1985yb} which treats theoretical uncertainties in a slightly more conservative way.} The branching fractions $\text{Br}(\Upsilon(nS)\rightarrow e e )$ can be taken from~\cite{Beringer:1900zz}.

\subsubsection*{Experimental constraints}
Turning to experiments, the BaBar collaboration has recently published several searches for light scalars in $\Upsilon$ decays. The results were presented in the form of upper limits on the product $\text{Br}(\Upsilon\rightarrow \gamma\,\phi)\times\text{Br}(\phi\rightarrow xx)$ with $xx$ being muons~\cite{Lees:2012iw}, taus~\cite{Lees:2012te}, gluons~\cite{Lees:2013vuj} and general hadronic final states~\cite{Lees:2011wb}. These can be translated into constraints on the coupling $y$  of the scalar $\phi$ to SM fields by using~\eqref{eq:upsilon} and the branching fractions from~\eqref{eq:branching}. The strongest bounds arise from $\tau\tau$ and hadronic final states; they are presented in Fig.~\ref{fig:blimits}.

\subsection{$B$ meson decays}

The scalar $\phi$ also gives rise to an effective flavor violating coupling $b-s-\phi$ which is obtained by integrating out the $W$-top-loop. One finds~\cite{Batell:2009jf}
\begin{equation}\label{eq:effcoupling}
\mathcal{L}_{\phi sb}=\frac{y\,m_b}{v}\,\frac{3\sqrt{2}\,  G_F \,m_t^2\, V_{ts}^* V_{tb}}{16\pi^2}\times \phi\, \bar{s}_L b_R +\text{h.c.} \;,
\end{equation}
with $V_{ts}$ and $ V_{tb}$ denoting the CKM elements. We follow~\cite{Freytsis:2009ct} and use the one-loop $\overline{\text{MS}}$ top mass $m_t=165\gev$ in the above expression.

For $m_\phi\lesssim 5\gev$, the scalar can mediate rare decays of $B$ mesons.
The most constraining mode is $B \rightarrow K+\phi$ for which the decay rate can be written as
\begin{align}
\small
\label{eq:BKphi}
&\Gamma^{B \rightarrow K\phi} = \left(\frac{y\,m_b}{v}\,\frac{3\sqrt{2}\,G_F\,m_t^2\,|V_{ts}^*V_{tb}|}{16\,\pi^2}\right)^2\left|\langle K|\bar{s}_L b_R|B\rangle\right|^2 \nonumber\\ &\quad \times \frac{\sqrt{(m_B^2-(m_K+m_\phi)^2)(m_B^2-(m_K-m_\phi)^2)}}{16\pi\,m_B^3}\;,
\end{align}
which agrees well with the numerical formula presented in~\cite{Batell:2009jf}. 
For the corresponding matrix element we use the parametrization~\cite{Ball:2004ye}
\begin{align}
 &\langle K|\bar{s}_L b_R|B\rangle = \frac{1}{2}\frac{(m_B^2-m_K^2)}{m_b-m_s}\,f_0(q^2) \nonumber \\ & \quad \text{with} \quad f_0(q^2)=\frac{0.33}{1-q^2/38\gev^2}\,,
\end{align}
with the transferred momentum $q^2=m_\phi^2$. This parametrization is in good agreement with a more recent determination of $f_0(q^2)$~\cite{Bouchard:2013mia}. The uncertainty of $f_0(q^2)$ is at the level of $\sim 10\%$~\cite{Ball:2004ye}. 

\subsubsection*{Experimental constraints}

The above decay mode would contribute to the rare process $B \rightarrow K + \mu\mu$ via $\phi$ decaying into a pair of muons (see right panel of Fig.~\ref{fig:penguin}).
As interference effects can be neglected -- the intermediate $\phi$ is on-shell -- this contribution simply adds to the SM one. The comparison with observation is still not straightforward as the experiments probe a regime of the coupling $y<0.01$, where the lifetime of $\phi$ becomes non-negligible (see Fig.~\ref{fig:branching}). If the scalar travels a macroscopic distance in the detector, this would affect the event reconstruction performed in the experimental analyses. Events with a too large displacement $\Delta d$ of the $\phi$-decay vertex from the primary interaction point would fail criteria on the vertex quality and be rejected as background. At LHCb $B$ mesons are produced with a higher boost than at $B$ factories. This typically leads to a larger displacement $\Delta d$ and to more events being rejected. Therefore the lower sensitivity of $B$ factories compared to LHCb is partially compensated as they miss less of the signal events. We hence consider the measurements of $B \rightarrow K 
+ \ell\ell$ at both, LHCb~\cite{Aaij:2012vr} and Belle~\cite{Wei:2009zv}.\footnote{BaBar has also performed a search for $B \rightarrow K + \ell\ell$ with sensitivity very similar to Belle~\cite{Lees:2012tva}.}  Note that $\ell=\mu$ at LHCb, while $\ell=\mu,e$ at Belle.

\begin{figure}[t]
\begin{center}  
  \includegraphics[width=6.0cm]{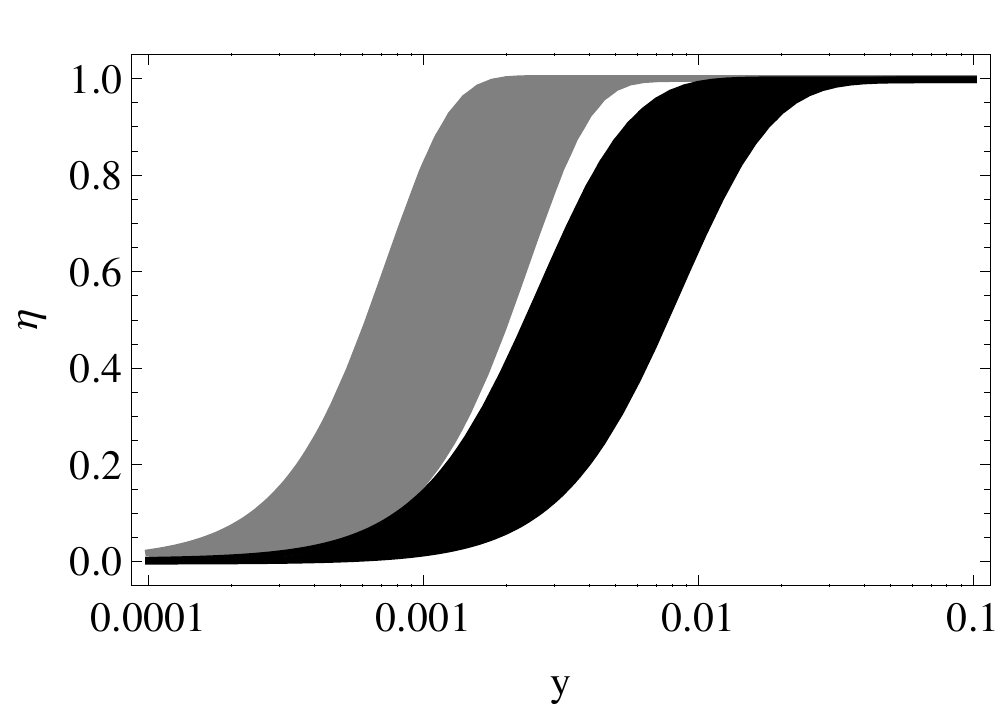}
\end{center}
\caption{Detection efficiency at LHCb (black) and Belle (gray) as a function of the light scalar coupling $y$ for fixed $m_\phi=1\gev$. The width of the bands arises from the uncertainty in the determination of $\eta$ (see text). For $y\gtrsim 0.1$ the efficiency approaches unity as $\phi$ decays promptly on detector scales. For $y\lesssim0.01$ the lifetime of $\phi$ becomes non-negligible leading to a displacement of the $\phi$ decay vertex. In this case, the efficiency decreases rapidly.}
\label{fig:efficiency}
\end{figure}

In order to constrain the coupling of $\phi$ to SM matter, we must take into account the efficiency $\eta$ of LHCb and Belle for reconstructing $\phi$-induced decays as a function of $m_\phi$ and $y$. Assuming that events with a displacement $\Delta d>d_{\text{max}}$ are rejected, we have
\begin{equation}
\label{eq:eff}
 \eta(m_\phi,y)=\int  \limits_0^{\infty} \!\!\D p_\phi \,f(p_\phi)\! \left[ 1- \exp\left(-\frac{m_\phi\,\Gamma_\phi\,d_\text{max}}{p_\phi}\right)\right]\!
\end{equation}
with $f(p_\phi)$ the momentum distribution of $\phi$. For LHCb we infer $f(p_\phi)$ from a large event sample generated with the Monte Carlo generator \PY~\cite{Sjostrand:2006za}, where we applied the appropriate pseudorapidity cuts in order to reject events which happen partially or fully outside the detector.\footnote{As a cross check we have determined the mean energy of the parent $B$ mesons passing the pseudorapidity cuts which we found to be around $100\gev$ depending slightly on $m_\phi$. This is in very good agreement with the value inferred from the more sophisticated LHCb detector simulation~\cite{Blake}.} In LHCb analyses with similar event selection as in~\cite{Aaij:2012vr} but with significant backgrounds from open charm decays, the vertex selection criteria remove a large fraction, but not all of the charmed mesons~\cite{Gershon}. This suggests a value of $d_{\text{max}}$ in the range $2\:\text{mm}-2\:\text{cm}$ which we have used in our analysis. 
As a cross check for our description of the detection efficiency of long-lived particles, we have determined $\eta$ for the Majorana neutrinos searches in~\cite{Aaij:2012zr}. Our prediction of $\eta$ using the indicated range of $d_{\text{max}}$ is in good agreement with efficiencies provided by the collaboration in Fig.~14 of~\cite{Aaij:2012zr}. The determination of the efficiency for Belle proceeds analogously. We take the same range $d_{\text{max}}=2\:\text{mm}-2\:\text{cm}$ as for LHCb. In order to verify that this assumption is justified, we have used the information from~\cite{Hyun:2010an}.\footnote{In~\cite{Hyun:2010an} Belle has performed a search for light scalars with subsequent decay to muons in the very narrow mass window $m_\phi=212-300\mev$. In the analysis, it was pointed out that the detection efficiency is only marginally affected for lifetimes $\Gamma_\phi^{-1}<10^{-12}\:\text{s}$. With our description of the efficiency and $d_{\text{max}}=2\:\text{mm}-2\:\text{cm}$ one would obtain $\eta=0.5-0.99$ for $\Gamma_\phi^{-1}=10^{-12}\:\text{s}$ in this example, showing that it is justified to use our range for $d_{\text{max}}$.} 
In Fig.~\ref{fig:efficiency}, we depict $\eta$ for LHCb and Belle. Due to the 
larger boosts at LHCb, the corresponding $\eta$ decreases more rapidly towards smaller $y$. 
For the example of $m_\phi=1\gev$, we see that in both experiments couplings $y\ll 10^{-3}$ cannot be constrained due to the loss in efficiency.

To set upper bounds on $y$, we employ the binned distribution of the branching ratio $B \rightarrow K + \ell\ell$ in the dilepton invariant mass $m_{\ell\ell}$. For $\phi$-induced decays $m_{\ell\ell}=m_\phi$, i.e.\ $m_\phi$ determines the bin in which the signal would occur. Using the SM distribution of $\text{Br}(B \rightarrow K+\mu\mu)$  from~\cite{Aaij:2012vr}\footnote{For the sake of a conservative approach, we use the lower end of the band provided in~\cite{Aaij:2012vr}. In the energy range $2.9-3.8 \gev$, where no SM prediction is given, we conservatively set it to zero.}, we calculate, bin by bin, the $90\%$ CL upper limit on the difference in branching compared to the SM. For this, we employ the Feldman-Cousins approach~\cite{Feldman:1997qc}. This limit is then translated into a constraint on the coupling $y$ by making use of \eqref{eq:branching}, \eqref{eq:BKphi} and \eqref{eq:eff}.

\section{Results and Summary}

\begin{figure}[t]
\begin{center}  
  \includegraphics[width=9cm]{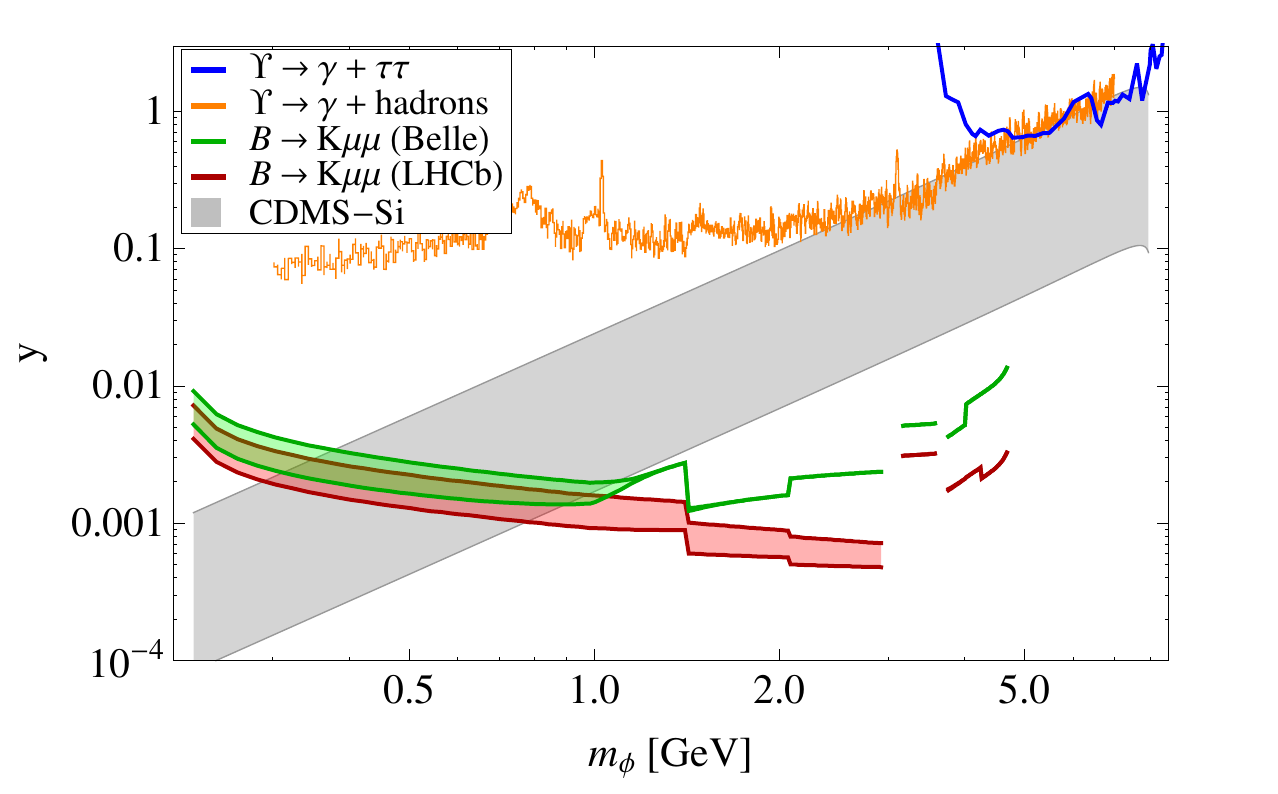}
\end{center}
\caption{Constraints on the coupling $y$ from $B$ and $\Upsilon$ decays.
The gray band indicates the range of DM-nucleon cross sections consistent with the CDMS-Si signal (for $m_\chi=8\gev$). The regions $m_\phi=2.95-3.18\gev$ and $m_\phi=3.59-3.77\gev$ are vetoed in the experimental searches for rare $B$ decays.}
\label{fig:blimits}
\end{figure}

In this letter, we have presented new severe constraints on light scalar particles with reduced Higgs-like couplings to the SM. For this we considered the most recent searches for rare $\Upsilon$ and $B$ decays performed at LHCb and $B$ factories. We carefully included detector effects related to the finite lifetime of the light scalar.

Our main results are summarized in Fig.~\ref{fig:blimits}, where we depict the resulting constraints on the reduced coupling $y$ of a light scalar $\phi$ to SM fields. An additional bound not shown in Fig.~\ref{fig:blimits} comes from LEP and constrains $y$ to be $y \lesssim 0.1$ for $m_\phi\leq 10\gev$~\cite{Acciarri:1996um}. The searches for radiative $\Upsilon$ decays have reached a similar sensitivity as LEP. If $m_\phi$ resides below the $B$ meson mass, considerably tighter constraints on $y$ are obtained from rare $B$ decays. We should emphasize that the limits shown in Fig.~\ref{fig:blimits} are valid independently of any assumptions about the dark sector, as long as the decay of $\phi$ into DM is kinematically forbidden. 

However, interesting bounds on the DM direct detection cross section can be obtained if $\phi$ mediates the interactions of DM with nuclei. After fixing the coupling between DM and $\phi$ by requiring the correct DM relic density, the cross section $\sigma_n$ of DM with nucleons is uniquely determined by the coupling $y$. In Fig.~\ref{fig:blimits}, we show the range in the coupling $y$, where $\sigma_n$ is consistent with the observed CDMS-Si signal. For illustration, we have fixed $m_\chi=8 \gev$, corresponding to the CDMS-Si favoured cross section $\sigma_n=4 \cdot 10^{-42}-8 \cdot 10^{-41} \text{cm}^2$ at $90 \%$ CL~\cite{Agnese:2013rvf}. It can be seen that the coupling required to obtain a sufficiently large $\sigma_n$ is excluded for $m_\phi\simeq 1-5\gev$ unless $m_\phi$ falls in the close vicinity of the charmed resonances. This conclusion is obtained for standard nuclear- and astrophysics assumptions. It is hardly affected if we vary the DM mass within the reasonable range $m_\chi=6-10 \gev$.\footnote{While the CDMS-Si result alone is consistent 
with $m_\chi>10\gev$, this possibility is safely excluded by other dark matter searches, in particular XENON -- even with the most conservative assumptions about experimental uncertainties~\cite{Aprile:2012nq}.}

We consider the remaining viable parameter region with mediator masses between $5\gev$ and $m_\chi$ particularly appealing as similar masses for both, DM and mediator,
can naturally arise in supersymmetric models~\cite{Kappl:2010qx}. 
Upcoming searches for radiative $\Upsilon$ decays with increased sensitivity could probe this intriguing region of parameter space. For mediator masses $m_\phi\lesssim 5\gev$, the sensitivity of the existing searches can already be significantly increased.
In particular LHCb already probes highly suppressed couplings of the mediator to SM fields, for which the scalar has a macroscopic decay length in the detector. By a modification of the event selection, the efficiency for such displaced decays could be optimized.

 
\section*{Acknowledgements}
We are indebted to Tim Gershon and Thomas Blake for sharing their invaluable insights into the LHCb vertex reconstruction. Further, we would like to thank Liang Sun and Kevin Flood for correspondence as well as Andreas Weiler for discussions. FS is supported by the BMBF PT DESY Verbundprojekt 05H2013-THEORIE `Vergleich von LHC-Daten mit supersymmetrischen Modellen'.


\bibliography{meson.bib}
\bibliographystyle{ArXiv}

\end{document}